\documentclass{llncs}

\newif\ifnotes
\notesfalse

\usepackage[T1]{fontenc}
\usepackage{amsmath}
\usepackage{txfonts}
\usepackage{xspace}
\usepackage{xcolor}
\usepackage{url}
\usepackage{verbatim}

\newcommand{\deq}{\mathrel{\smash{\stackrel{\scriptscriptstyle\mathrm{\Delta}}{=}}}}
\renewcommand{\triangleq}{\deq}
\renewcommand\times{*}

\newcommand{\abbr}[1]{%
\ifmmode\mathrm{\MakeUppercase{#1}}\else{\upshape\rmfamily\MakeUppercase{#1}}\fi}
\newcommand{\tlaplus}{\ensuremath{\abbr{TLA}\kern-.2ex{}^+}\xspace}
\newcommand{\tlaplustwo}{\ensuremath{\abbr{TLA}\kern-.2ex{}^{+2}}}
\newcommand{\tlaps}{\abbr{TLAPS}\xspace}
\newcommand{\tlapm}{\abbr{TLAPM}\xspace}
\newcommand{\isatlaplus}{Isabelle/\tlaplus}
\newcommand{\kwd}[1]{\ensuremath{\text{\textsc{#1}}}}

\newcommand{\TRUE}{\textsc{true}}

\definecolor{brown}{rgb}{0.4,0.2,0.1}
\ifnotes
  \newenvironment{ednote}[2]{\begin{quote}\sf\footnotesize\color{#1}#2: }{\end{quote}}
  \newcommand{\mgnote}[3]{\marginpar{\footnotesize\textcolor{#1}{\sf #2: #3}}}
  \newenvironment{kc}{\begin{ednote}{blue}{KC}}{\end{ednote}}
  \newenvironment{dd}{\begin{ednote}{green}{DD}}{\end{ednote}}
  \renewenvironment{ll}{\begin{ednote}{red}{LL}}{\end{ednote}}
  \newenvironment{sm}{\begin{ednote}{brown}{SM}}{\end{ednote}}
\else

  \newcommand{\mgnote}[3]{}
\fi

\title{Verifying Safety Properties With the \tlaplus Proof System}
\author{
  Kaustuv Chaudhuri\inst{1}
  \and
  Damien Doligez\inst{2}
  \and
  Leslie Lamport\inst{3}
  \and
  Stephan Merz\inst{4}}
\institute{
  INRIA Saclay, France, \email{kaustuv.chaudhuri@inria.fr}
  \and
  INRIA Rocquencourt, France, \email{damien.doligez@inria.fr}
  \and
  Microsoft Research Silicon Valley, USA, \email{http://lamport.org}
  \and
  INRIA Nancy, France, \email{stephan.merz@inria.fr}
}

\begin{document}

\maketitle

\thispagestyle{plain}
\pagestyle{plain}

\section{Overview}
\label{sec:over}

\tlaps, the \tlaplus proof system, is a platform for the development and
mechanical verification of \tlaplus proofs. The \tlaplus proof language is
declarative, and understanding proofs requires little background beyond
elementary mathematics. The language supports hierarchical and non-linear proof
construction and verification, and it is independent of any verification tool or
strategy. Proofs are written in the same language as specifications; engineers
do not have to translate their high-level designs into the language of a
particular verification tool. A \emph{proof manager} interprets a \tlaplus
proof as a collection of \emph{proof obligations} to be verified, which it sends
to \emph{backend verifiers} that include theorem provers, proof assistants, SMT
solvers, and decision procedures.

The first public release of \tlaps is available from~\cite{tlaps-website},
distributed with a BSD-like license. It handles almost all the non-temporal part
of \tlaplus as well as the temporal reasoning needed to prove standard safety
properties, in particular invariance and step simulation, but not liveness
properties. Intuitively, a safety property asserts what is permitted to happen;
a liveness property asserts what must happen; for a more formal overview,
see~\cite{alpern85ipl,lamport77ieeese}.

\section{Foundations}
\label{sec:found}

\tlaplus is a formal language based on TLA (the Temporal Logic of
Actions)~\cite{lamport03tlabook}. It was designed for specifying the high-level
behavior of concurrent and distributed systems, but it can be used to specify
safety and liveness properties of any discrete system or algorithm. A behavior
is a sequence of states, where a state is an assignment of values to \emph{state
  variables}. Safety properties are expressed by describing the allowed steps
(state transitions) in terms of \emph{actions}, which are first-order formulas
involving two copies $v$ and $v'$ of each state variable, where $v$ denotes the
value of the variable at the \emph{current} state and $v'$ its value at the
\emph{next} state. These properties are proved by reasoning about actions, using
a small and restricted amount of temporal reasoning. Proving liveness properties
requires propositional linear-time temporal logic reasoning plus a few TLA proof
rules.

It has always been possible to assert correctness properties of systems in \tlaplus,
but not to write their proofs. We have added proof constructs based on a
hierarchical style for writing informal proofs~\cite{lamport95amm}. The current
version of the language is essentially the same as the version described
elsewhere~\cite{chaudhuri08keappa}. Here, we describe only the \tlaps proof
system.
Hierarchical proofs are a stylistic variant of natural deduction with lemmas
and have been used in other declarative proof
languages~\cite{corbineau07types,rudnicki92types,wenzel09isar}. A hierarchical
proof is either a sequence of steps together with their proofs, or a leaf
(lowest-level) proof that simply states the known facts (previous steps and
theorems) and definitions from which the desired conclusion follows. The human
reader or a backend verifier must ensure that the leaf proofs are correct in
their interpretation of \tlaplus to believe the entire proof.

The \tlaps proof manager, \tlapm, reads a (possibly incomplete) hierarchical
proof and invokes the backend verifiers to verify the leaf proofs. One important
backend is \isatlaplus, which is an implementation of \tlaplus as an Isabelle
object logic (see Section~\ref{sec:backends.isa}). \isatlaplus can be used directly
with Isabelle's generic proof methods,
or other certifying backend verifiers can produce proofs that are
checked by \isatlaplus. Currently, the only certifying backend is the Zenon
theorem prover~\cite{bonichon07lpar}. Among the non-certifying backends is a
generic SMT-LIB-based backend for SMT solvers, and a decision procedure for
Presburger arithmetic. We plan to replace these with
certifying implementations such as the SMT solver veriT~\cite{bouton09cade} and
certifying implementations of decision procedures~\cite{chaieb08jar}.

\tlaps is intended for avoiding high-level errors in systems, not for
providing a formal foundation for mathematics. It is far more likely for a
system error to be caused by an incomplete or incorrect specification than by an
incorrect proof inadvertently accepted as correct due to bugs in \tlaps.
Although we prefer certifying backends whenever possible, we include
non-certifying backends for automated reasoning in important theories such as
arithmetic.

\section{Proof management}
\label{sec:obgen}

A \tlaplus specification consists of a root module that can (transitively)
import other modules by extension and parametric instantiation. Each module
consists of a number of parameters (state variables and uninterpreted
constants), definitions, and theorems that may have proofs. \tlaps is run by
invoking the Proof Manager (\tlapm) on the root module and telling it which
proofs to check. In the current version, we use pragmas to indicate the proofs
that are not to be checked, but this will change when \tlaps is integrated into
the \tlaplus Toolbox IDE~\cite{tla-toolbox}. The design of \tlapm for the
simple constant expressions of \tlaplus was described
in~\cite{chaudhuri08keappa}; this section explains the further processing
required to support more of the features of \tlaplus. \tlapm first flattens the
module structure, since the module language of \tlaplus is not supported by
backend verifiers, which will likely remain so in the future.

\paragraph{Non-constant reasoning:}

A \tlaplus module parameter is either a \emph{constant} or a (state)
\emph{variable}. Constants are independent of behaviors and have the same value
in each state of the behavior, while a variable can have different values in
different states. Following the tradition of modal and temporal logics, \tlaplus
formulas do not explicitly refer to states. Instead, action formulas are built
from two copies $v$ and $v'$ of variables that refer to the values before and
after the transition. More generally, the prime operator $'$ can be applied to
an entire expression $e$, with $e'$ representing the value of $e$ at the state
after a step. A \emph{constant expression} $e$ is one that does not involve any
state variables, and is therefore equal to $e'$. (Double priming is not allowed
in \tlaplus; the \tlaplus syntactic analyzer catches such errors.)

Currently, all \tlaps backends support logical reasoning only on constant
expressions. The semantics of the prime operator is therefore syntactically
approximated as follows: it is commuted with all ordinary operators of
mathematics and is absorbed by constant parameters. Thus, if $e$ is the
expression $(u = v + 2 \times c)$ where $u$ and $v$ are variables and $c$ a
constant, then $e'$ equals $u' = v' + 2 \times c$. \tlapm currently performs
such rewrites and its rewrite engine is trusted.

\paragraph{Operators and substitutivity:}

At any point in the scope of its definition, a user-defined operator is in one
of two states: \emph{usable} or \emph{hidden}. A usable operator is one whose
definition may be \emph{expanded} in a proof; for example, if the operator $P$
defined by $P(x, y) \triangleq x + 2 \times y$ is usable, then \tlapm may
replace $P(2, 20)$ with $2 + 2 \times 20$ (but \emph{not} with 42, which
requires proving that $2 + 2 \times 20 = 42$). A user-defined operator is hidden
by default; it is made usable in a particular leaf proof by explicitly citing
its definition, or for the rest of the current subproof by a \kwd{use} step
(see~\cite{chaudhuri08keappa} for the semantics of \kwd{use}).

Because \tlaplus is a modal logic, it contains operators that do not obey
substitutivity, which underlies Leibniz's principle of equality. For example,
from $(u = 42) = \TRUE$ one cannot deduce $(u=42)' = \TRUE'$, \emph{i.e.}, $u'=42$. A
unary operator $O(\_)$ is \emph{substitutive} if $e=f$ implies $O(e)=O(f)$, for
all expressions $e$ and $f$. This definition is extended in the obvious way to
operators with multiple arguments. Most of the modal primitive operators of
\tlaplus are not substitutive; and an operator defined in terms of
non-substitutive operators can be non-substitutive. If a non-substitutive
operator is usable, then \tlapm expands its definition during preprocessing, as
described in the previous paragraph; if it is hidden, then \tlapm replaces its
applications by cryptographic hashes of its text to prevent unsound inferences
by backend verifiers. This is a conservative approximation: for example, it
prevents proving $O(e \land f) = O(f \land e)$ for a hidden non-substitutive
operator $O$. Users rarely define non-substitutive operators, so there seems to
be no urgent need for a more sophisticated treatment.

\paragraph{Subexpression references:}

A fairly novel feature of the \tlaplus proof language is the ability to refer to
arbitrary subexpressions and instances of operators, theorems, and proof steps
that appear earlier in the module or in imported modules, reducing the verbosity
and increasing the maintainability of \tlaplus proofs. \emph{Positional}
references denote a path through the abstract syntax; for example, for the
definition, $\mathit{O}(x, y) \deq x = 20 * y + 2$, the reference
$\mathit{O}(3,4)!2!1$ resolves to the first subexpression of the second
subexpression of $\mathit{O}(3,4)$, \emph{i.e.}, $20 * 4$. Subexpressions can
also be labelled and accessed via \emph{labelled} references. For example, for
$\mathit{O}(x, y) \deq x = l{::}(y * 20) + 2$, the reference $\mathit{O}(3,4)!l$
refers to $4 * 20$ and will continue to refer to this expression even if the
definition of $\mathit{O}$ is later modified to $\mathit{O}(x, y) \deq x = 7 *
y^2 + l{::}(20 * y) + 2$. \tlapm replaces all subexpression references with the
expressions they resolve to prior to further processing.

\paragraph{Verifying obligations:}

Once an obligation is produced and processed as described before, \tlapm invokes
backend verifiers on the proof obligations corresponding to the leaf proofs. The
default procedure is to invoke the Zenon theorem prover first. If Zenon succeeds
in verifying the obligation, it produces an Isabelle/Isar proof script that can
be checked by \isatlaplus. If Zenon fails to prove an obligation, then
\isatlaplus is instructed to use one of its automated proof methods. The default
procedure can be modified through pragmas that instruct \tlapm to bypass Zenon,
use particular Isabelle tactics, or use other backends. Most users will invoke
the pragmas indirectly by using particular theorems from the standard
\url{TLAPS} module. For instance, using the theorem named
\texttt{SimpleArithmetic} in a leaf proof causes \tlapm to invoke a decision
procedure for Presburger arithmetic for that proof. The user can learn what
standard theorems can prove what kinds of assertions by reading the
documentation, but she does not need to know how such standard theorems are
interpreted by \tlapm.

\section{Backend verifiers}
\label{sec:backends}

\subsection{\isatlaplus}
\label{sec:backends.isa}

\isatlaplus is an axiomatization of \tlaplus in the generic proof assistant
Isabelle~\cite{paulson94isabelle}. It embodies the semantics of the constant
fragment of \tlaplus in \tlaps; as mentioned in Section~\ref{sec:found}, it is
used to certify proofs found by automatic backend verifiers. We initially
considered encoding \tlaplus in one of the existing object logics that come with
the Isabelle distribution, such as Isabelle/ZF or Isabelle/HOL. However, this
turned out to be inconvenient, mainly because \tlaplus is untyped. (Indeed,
\tlaplus does not even distinguish between propositions and terms.) We would
have had to define a type of \tlaplus values inside an existing object
logic and build \tlaplus-specific theories for sets, functions, arithmetic
\emph{etc.}, essentially precluding reuse of the existing infrastructure.

\isatlaplus defines classical first-order logic based on equality, conditionals,
and Hilbert's choice operator. All operators take arguments and return values of
the single type \texttt{c} representing \tlaplus values. Set theory is based on
the uninterpreted predicate symbol $\in$ and standard Zermelo-Fr\"ankel axioms.
Unlike most presentations of ZF, \tlaplus considers functions to be primitive
objects rather than sets of ordered pairs. Natural numbers with zero and
successor are introduced using Hilbert's choice as some set satisfying the Peano
axioms; the existence of such a set is established from the ZF axioms. Basic
arithmetic operators over natural numbers such as $\leq$, $+$, and $*$ are
defined by primitive recursion, and division and modulus are defined in terms of
$+$ and~$*$. Tuples and sequences are defined as functions whose domains are
initial intervals of the natural numbers. Characters are introduced as pairs of
hexadecimal digits, and strings as sequences of characters. Records are
functions whose domains are finite sets of strings. Isabelle's flexible parser
and pretty-printer transparently converts between the surface syntax and the
internal representation. The standard library introduces basic operations for
these data structures and proves elementary lemmas about them. It currently
provides more than 1400 lemmas and theorems, corresponding to about 200 pages of
pretty-printed Isar text. \isatlaplus sets up Isabelle's generic automated proof
methods (rewriting, tableau and resolution provers, and their combinations).

It is a testimony to the genericity of Isabelle that setting up a new object
logic was mostly a matter of perseverance and engineering. Because \tlaplus is
untyped, many theorems come with hypotheses that express ``typing conditions''.
For example, proving $n+0 = n$ requires proving that $n$ is a number. When the
semantics of \tlaplus allowed us to do so, we set up operators so that they
return the expected ``type''; for example, $p \land q$ is guaranteed to be a
Boolean value whatever its arguments $p$ and $q$ are. In other cases,
typechecking is left to Isabelle's automatic proof methods; support for
conditional rewrite rules in Isabelle's simplifier was essential to make this
work.

\subsection{Zenon}
\label{sec:backends.zenon}

Zenon is a theorem prover for first-order logic with Hilbert's choice operator
and equality. It is a \emph{proof-producing} theorem prover: it outputs formal
proof scripts for the theorems it proves. Zenon was extended with a backend that
produces proofs in Isar syntax; these proofs use lemmas based on the \isatlaplus
object logic and are passed to Isabelle for verification. Zenon is
therefore not part of the trusted code base of \tlaps.

Zenon had to be extended with deduction rules specific to \tlaplus: rules for
reasoning about set-theoretic operators, for the \kwd{case} operator of
\tlaplus, for set extensionality and function extensionality, for reasoning
directly on bounded quantifiers (which is not needed in theory but is quite
important for efficiency), and for reasoning about functions, strings,
\emph{etc}. Interestingly, Hilbert's choice operator was already used in Zenon
for Skolemization, so we were easily able to support the \kwd{choose} operator of
\tlaplus.

Future work includes adding rules to deal with tuples, sequences, records, and
arithmetic, and improving the handling of equality. While there is some overlap
between Zenon and Isabelle's automatic methods as they are instantiated in
\isatlaplus, in practice they have different strong points and there are many
obligations where one succeeds while the other fails.
Zenon uses Isabelle's automatic proof tactics for some of the elementary steps
when it knows they will succeed, in effect using these tactics as high-level
inference rules.

\subsection{Other backends}
\label{sec:backends.other}

The first release of \tlaps comes with some additional non-certifying
backends. For arithmetic reasoning we have:
\begin{itemize}
\item An SMT-LIB based backend that can be linked to any SMT solver.
  Obligations are rewritten into the AUFLIRA theory of SMT-LIB, which generally
  requires omitting assumptions that lie outside this theory. This backend is
  needed for reasoning about real numbers. We have successfully used Yices,
  CVC3, Z3, veriT and Alt-Ergo in our examples. In future work we might
  specialize this generic backend for particular solvers that can reason about
  larger theories.
\item A Presburger arithmetic backend, for which we have implemented Cooper's
  algorithm. Our implementation is tailored to certain elements of \tlaplus that
  are not normally part of the Presburger fragment, but can be (conservatively)
  injected.
\end{itemize}
For both these backends, \tlapm performs a simple and highly conservative sort
detection pass for bound identifiers. Both backends are currently
non-certifying, but we plan to replace them with certifying backends in the
future. In particular, we are integrating the proof-producing SMT solver
veriT~\cite{bouton09cade}, with the goal of tailoring it for discharging
\tlaplus proof obligations.

\section{Proof development}
\label{sec:dev}

Writing proofs is hard and error-prone. Before attempting to prove correctness
of a \tlaplus specification, we first check finite instances with the TLC model
checker~\cite{lamport03tlabook}. This usually catches numerous errors quickly --
much more quickly than by trying to prove it correct. Only after TLC can find no
more errors do we try to write a proof.

The \tlaplus language supports a hierarchical, non-linear proof development
process that we find indispensable for larger proofs~\cite{gafni03disc}. The
highest-level proof steps are derived almost without thinking from the structure
of the theorem to be proved. For example, a step of the form $P_{1} \lor \ldots
\lor P_{n} \Rightarrow Q$ is proved by the sequence of steps asserting
$P_{i}\Rightarrow Q$, for each~$i$. When the user reaches a simple enough step,
she first tries a fully automatic proof using a leaf directive citing the facts
and definitions that appear relevant. If that fails, she begins a new level with
a sequence of proof-less assertion steps that simplify the assertion, and a
final \kwd{qed} step asserting that the goal follows from these steps. These new
lower-level steps are tuned until the \kwd{qed} step is successfully verified.
Then, the steps are proved in any order. (The user can ask \tlapm what steps
have no proofs.)
The most common reason that leaf proofs fail to verify is that the user
has forgotten to use some fact or definition. When a proof fails, \tlapm prints
the usable hypotheses and the goal, with usable definitions expanded. Examining
this output often reveals the omission.

This kind of hierarchical development cries for a user interface that allows one
to see what has been proved, hide irrelevant parts of the proof, and easily tell
\tlapm what it should try to prove next. Eventually, these functions will be
provided by the \tlaplus Toolbox. (It now performs only the hiding.) When \tlaps
is integrated into the Toolbox, writing the specification, model-checking it,
and writing a proof will be one seamless process. Meanwhile, we have written an
Emacs mode that allows hierarchical viewing of proofs and choosing which parts
to prove.

We expect most users to assume simple facts about data structures such as
sequences rather than spending time proving them -- especially at the beginning,
before we have developed libraries of such facts for common data structures.
Relying on unchecked assumptions would be a likely source of errors; it is easy
to make a mistake when writing an ``obviously true'' assumption. Such
assumptions should therefore be model-checked with TLC.

\subsection{Example developments}
\label{sec:exp}

We have written a number of proofs, mainly to find bugs and see how well the
prover works.  Most of them are in the \texttt{examples} sub-directory of the
\tlaps distribution.  Here are the most noteworthy:
\begin{itemize}
\item \emph{Peterson's Mutual Exclusion Algorithm.} This is a standard shared
  memory mutual exclusion algorithm. The algorithm (in its 2-process version) is
  described in a dozen lines of PlusCal, an algorithm language that is
  automatically translated to \tlaplus. The proof of mutual exclusion is about
  130 lines long.
\item \emph{The Bakery Algorithm with Atomic Reads and Writes.} This is a more
  complicated standard mutual exclusion example; its proof (for the $N$-process
  version) is 800 lines long.
\item \emph{Paxos.} We have specified a high-level version of the well-known
  Paxos consensus algorithm as a trivial specification of consensus and two
  refinement steps---a total of 100 lines of \tlaplus. We have completed the
  proof of the first refinement and most of the proof of the second. The first
  refinement proof is 550 lines long; we estimate that the second will be
  somewhat over 1000 lines.
\end{itemize}
Tuning the back-end provers has made them more powerful, making proofs easier to
write. 
While writing machine-checked proofs remains tiresome and more time
consuming than we would like, it has not turned out to be difficult once the
proof idea has been understood.

\paragraph{Acknowledgements} Georges Gonthier helped design the \tlaplus proof
language. Jean-Baptiste Tristan wrote the (incomplete) Paxos proof.

\bibliographystyle{abbrv}
\bibliography{tlaps}

\end{document}